\documentclass[pteplogo]{ptephy_v2}

\preprintnumber{XXXX-XXXX} 
\usepackage{hyperref}
\usepackage{color}

\usepackage{amsmath} 
\usepackage{mathtools}
\usepackage{subfig} 
\usepackage[version=3]{mhchem}

\newcommand{\nbar}{{\bar{n}}}
\newcommand{\nnbar}{{n\text{--}\nbar}}
\begin{document}

\title{Impact of a Reflecting Material on a Search for Neutron--Antineutron Oscillations using Ultracold Neutrons}

\author{Hiroyuki Fujioka}
\affil{Department of Physics, Institute of Science Tokyo, 2-12-1 Ookayama, Meguro, Tokyo 152-8551, Japan\email{fujioka@phys.sci.isct.ac.jp}}

\author[2,3]{Takashi Higuchi}
\affil{Institute for Integrated Radiation and Nuclear Science, Kyoto University, 2 Asashiro-nishi, Kumatori-cho, Osaka 590-0494, Japan}

\affil[3]{Research Center for Nuclear Physics, The University of Osaka,10-1 Mihogaoka, Ibaraki, Osaka 567-0047, Japan}


\begin{abstract}%
We investigate neutron--antineutron oscillations of ultracold neutrons in a storage bottle represented by a one-dimensional potential. The experimental sensitivity is determined by the annihilation rate of antineutrons. Its dependence on the antineutron reflectivity and the relative phase shift between the neutron and the antineutron wavefunctions by a reflection from the wall is derived. 
Optimization of the antineutron pseudopotential was found crucial to maximize the sensitivity of the experiment.
Furthermore, methods are discussed for determining the antineutron pseudopotential, which has only been studied indirectly thus far.
\end{abstract}

\subjectindex{C08,C12}

\maketitle

\section{Introduction}
Neutron--antineutron ($\nnbar$) oscillations, which violate both $\mathcal{B}$ and $\mathcal{B}-\mathcal{L}$ ($\mathcal{B}$: baryon number, $\mathcal{L}$: lepton number), are predicted in models beyond the Standard Model~\cite{Phillips2016-ll}.
The methods of experimental searches are categorized into two: i) $\nnbar$ oscillations of free neutrons and ii) oscillations of neutrons bound in nuclei.
The current limits on the $\nnbar$ oscillation time are $\tau_{n-\bar{n}}>0.86\times 10^8\,\mathrm{sec}$ (90\% C.L.) with free neutrons at the Institut Laue-Langevin (ILL)~\cite{Baldo-Ceolin1994-gj}, and $\tau_{n-\bar{n}}>4.7\times 10^8\,\mathrm{sec}$ (90\% C.L.) for bound neutrons obtained from the lower limit of $\ce{^16O}$ half-life at Super-Kamiokande~\cite{Abe2021-td}.
Further improvements in experimental sensitivity are required to probe theoretically motivated oscillation times; for example, an oscillation time as large as $5\times 10^{10}\,\mathrm{sec}$ in post-sphaleron baryogenesis models was predicted~\cite{Babu2013-mb}.
The HIBEAM/NNBAR experiment at the European Spallation Source aims to improve the limits with free neutrons~\cite{Addazi2021-ts,Santoro2024}, while Hyper-Kamiokande experiment and the Deep Underground Neutrino Experiment~\cite{Barrow2020-hg,Abi2021-mi} plan to update the half-life limits of $\ce{^{16}O}$ and $\ce{^{40}Ar}$, which can be converted into limits on the $\nnbar$ oscillation time for bound neutrons.

While the first two approaches have been demonstrated experimentally, a third approach employing ultracold neutrons (UCNs) has so far remained unrealized despite extensive discussion in Refs.~\cite{Chetyrkin1981-wk,Golub1989-qu,Yoshiki1992-eh,Kerbikov2003-cw,Kerbikov2004,Shima2025-ie}. UCNs are neutrons with kinetic energies on the order of $100\,\mathrm{neV}$ or less, which can be confined for times on the order of $100\,\mathrm{sec}$ by a material potential. This presents a notable advantage for $\nnbar$ oscillation searches through their long observational times, enabling experiments with a more compact setup than the beam experiments~\cite{Baldo-Ceolin1994-gj,Santoro2024} (e.g. $200\,\mathrm{m}$ in the case of the HIBEAM/NNBAR experiment).

Furthermore, recent developments in this field have brought about the advent of so-called superthermal UCN sources, which utilize inelastic scattering of neutrons from phonons excited in suitable media~\cite{Golub1977-kg,Masuda2012-gg,Ito2018-aa,Bison2020,Degenkolb2025-at,Algohi2025}. These sources enable the production of UCNs with densities that are orders of magnitude higher than those of conventional sources. This offers a substantial gain for fundamental physics experiments with UCNs~\cite{Abel2020-xe,Gonzalez2021-sc}.

These advances motivate a reexamination of the approach to $\nnbar$ oscillation searches using stored UCNs and a discussion of its potential reach with the state-of-the-art UCN technologies.
In this paper, we study an experiment to search for the $\nnbar$ oscillations using stored UCNs within a one-dimensional framework, focusing on the effects of the reflections from the wall on the experimental sensitivity.
The novelty of this paper is that we adopt the least approximation to derive a simple expression of the antineutron wavefunction which is applicable to realistic experimental conditions.

This paper is organized as follows.
Section~\ref{sec:UCNreflection} introduces UCN reflection by a material potential. 
In Section~\ref{sec:nnbar_freespace}, we briefly review a general framework of $\nnbar$ oscillations in free space. 
Next, in Section~\ref{sec:nnbar_in_bottle} we derive the antineutron wavefunction and discuss a measure of the experimental sensitivity.
Based on these results, we consider the optimum conditions for the reflecting material of a UCN storage bottle to maximize the experimental sensitivity in Section~\ref{sec:UCNmaterial}.
In Section~\ref{sec:scatteringlength}, we discuss experimental ideas using antiprotons and antineutrons to provide important inputs for optimizing the material.
Finally, we conclude this paper in Section~\ref{sec:conclusion}.

\section{Reflection of UCNs from a wall}\label{sec:UCNreflection}
Slow neutrons experience an effective optical potential in a material, known as Fermi's pseudopotential. It is expressed as $U_n=(2\pi \hbar^2/m)N[(A+1)/A]a_{nA}$, where $m$ is the neutron mass, $N$ is the number density of atoms, $A$ is the mass number, and $a_{nA}$ is the neutron--nucleus ($nA$) scattering length.
The real part of the $U_n$ are typically on the order of 10 to $100\,\mathrm{neV}$, while the imaginary part are 4 to 6 orders of magnitudes smaller than the real part with a few exceptions.
Hence, UCNs satisfying $E<U_n$ are totally reflected by the material surface.

In the following sections, we will discuss UCNs contained in a UCN storage bottle by its pseudopotential $V_0-iW_0$. We consider UCNs with kinetic energy $E<V_0$ that can be trapped in the bottle.
For later use, we briefly outline the one-dimensional motion of a neutron with a kinetic energy $E$ in a step potential $U(x)$:
\begin{align}
    U(x)=
    \begin{cases}
        0 & (x<0)\\
        V_0-iW_0 & (x>0)
    \end{cases}.
\end{align}
Solving the Schr\"{o}dinger equation, we obtain 
the wavefunction of a neutron as
\begin{align}
    \psi(x,t)=
    \begin{cases}
        \displaystyle\exp\left(ikx-\frac{iEt}{\hbar}\right)+\frac{k-k'}{k+k'}\exp\left(-ikx-\frac{iEt}{\hbar}\right) & (x<0) \\
        \displaystyle\frac{2k}{k+k'}\exp\left(ikx-\frac{iEt}{\hbar}\right) & (x>0)
    \end{cases},
\end{align}
with $k=\sqrt{2mE}/\hbar$ and $k'=\sqrt{2m[E-(V_0-iW_0)]}/\hbar$.
The reflection coefficient, $r=(k-k')/(k+k')$, defined as the ratio of reflected to incident wave amplitudes, gives the reflectivity as $|r|^2$.
In particular, for a real potential ($W_0=0$), the reflectivity is unity, as $k'$ is purely imaginary.
In addition, the reflected wave acquires a phase shift $\arg[(k-k')/(k+k')]$, while the wavefunction for $x>0$ decays exponentially since $\mathrm{Im}\,(k')>0$.

In what follows, we extend this pseudopotential treatment to antineutrons and discuss a setup for an $\nnbar$ oscillation search experiment with stored UCNs. One basis for this extension is that the pseudopotential formalism has been experimentally validated through neutron reflection from absorptive materials such as $^{113}$Cd and $^{155}$Gd~\cite{PismaZhETF.46.301}.

Even though it is more accurate to represent the neutron motion by a wave packet,
a semi-classical treatment using a plane wave will be adopted in the following sections, as was done in Refs.~\cite{Golub1989-qu,Kerbikov2004}.

\section{Neutron--antineutron oscillation in free space}\label{sec:nnbar_freespace}
First, we outline the formalism of $\nnbar$ oscillations in free space with a uniform magnetic field $B$~\cite{Phillips2016-ll}.
For the sake of brevity, we ignore the neutron decay; the exact form of the antineutron appearance probability includes an additional factor of $\exp(-t/\tau_n)$, where $\tau_n$ is the neutron lifetime. 

A two-component wavefunction
\begin{align}
\Psi(t)\exp(\pm ikx)=\begin{pmatrix} \psi_n(t)\exp(\pm ikx) \\ \psi_{\bar{n}}(t)\exp(\pm ikx)\end{pmatrix}
\end{align}
satisfies the following Schr\"{o}dinger equation:
\begin{align}
i\hbar\frac{d}{dt}\Psi(t)=\mathcal{H}\Psi(t),   
\end{align}
with an effective Hamiltonian $\mathcal{H}$ represented by
\begin{align}
    \mathcal{H}=\begin{pmatrix}
    E-\Delta E & \delta m\\
    \delta m & E+\Delta E
    \end{pmatrix}.
\end{align}
The neutron magnetic moment $\mu_n$ causes an energy splitting of $\mp \Delta E=\pm \mu_n B$.
The off-diagonal term $\delta m$ represents the transition amplitude of $\nnbar$ oscillation and is real under CP symmetry.
In the absence of a magnetic field, the antineutron appearance probability is
\begin{align}
    P_\nbar(t)=|\psi_\nbar(t)|^2=\sin^2\left(\frac{\delta m\,t}{\hbar}\right)=\sin^2\left( \frac{t}{\tau_\nnbar} \right),
\end{align}
where $\tau_\nnbar\equiv \hbar/\delta m$ is called the oscillation time.
The current lower limits of the oscillation time suggest $\delta m<10^{-29}\,\mathrm{MeV}$.
Even though the absence of a magnetic field would ideally minimize the energy splitting, residual fields cannot be completely eliminated in practice. As a result, $\Delta E=|\mu_n| B=6.03\times 10^{-23}\,\mathrm{MeV}\times (B/1\,\mathrm{nT})$ remains orders of magnitude larger than $\delta m$. By diagonalizing the Hamiltonian, the time evolutions of $\psi_n(t)$ and $\psi_\nbar(t)$ are obtained. The results are
\begin{align}
    \psi_n(t)&=\left[\cos\frac{\Delta E't}{\hbar}+i\cos\theta\sin\frac{\Delta E't}{\hbar}\right]\exp\left(-\frac{iEt}{\hbar}\right),\\
    \psi_\nbar(t)&=-i\sin\theta\sin\frac{\Delta E't}{\hbar}\exp\left(-\frac{iEt}{\hbar}\right),\label{psi_nbar}
\end{align}
where $\Delta E'=\sqrt{(\Delta E)^2+(\delta m)^2}$ and $\theta=\tan^{-1}(\delta m/\Delta E)$.

The same result is reached by an approximation of 
\begin{align}
    \psi_n(t)=\exp\left[-\frac{i(E-\Delta E)t}{\hbar}\right],\label{common_psi_n}
\end{align}
which is justified since $1-\cos\theta\approx \theta^2/2$ is always very small.
Indeed, the Schr\"{o}dinger equation for $\psi_\nbar(t)$:
\begin{align}
    i\hbar\frac{d}{dt}\psi_\nbar(t)=\delta m\exp\left[-\frac{i(E-\Delta E)t}{\hbar}\right]+(E+\Delta E)\psi_\nbar(t)\label{eq:Schrodinger}
\end{align}
has a solution of
\begin{align}
    \psi_\nbar(t)&=-\frac{\delta m}{2\Delta E}\exp\left[-\frac{i(E-\Delta E)t}{\hbar}\right]+\frac{\delta m}{2\Delta E}\exp\left[-\frac{i(E+\Delta E)t}{\hbar}\right]\label{eq:psi_nbar_sol}\\
    &=-i\frac{\delta m}{\Delta E}\sin\frac{\Delta E\,t}{\hbar}\exp\left(-\frac{iEt}{\hbar}\right),
\end{align}
and it is essentially the same as Eq.~(\ref{psi_nbar}),
as $\delta m/\Delta E\approx \tan\theta \approx \sin \theta$ and $\Delta E\approx \Delta E'$.
In a later discussion of the time evolution of $\bar{\psi}_n$ in a UCN storage bottle, we adopt this approximation to render the problem more tractable.

It is noted that as long as $\Delta E\,t\ll \hbar$ is satisfied, the antineutron appearance probability is approximated by
\begin{align}
    P_\nbar(t)=|\psi_\nbar(t)|^2\approx \left(\frac{\delta m}{\Delta E}\right)^2 \left(\frac{\Delta E\,t}{\hbar}\right)^2=\frac{t^2}{\tau_\nnbar^2},\label{nbar_appearance_prob}      
\end{align}
which grows quadratically with the elapsed time. This condition is  called the quasi-free condition.

\section{Neutron--antineutron oscillation in a storage bottle}\label{sec:nnbar_in_bottle}

\subsection{Theoretical framework of the problem}
Let us consider a UCN with a kinetic energy $E$ confined in one-dimensional potential $U_n(x)$ due to the material of the storage bottle:
\begin{align}
    U_n(x)=
    \begin{cases}
        V_0 & (x<-\ell/2)\\
        0 & (-\ell/2<x<\ell/2)\\
        V_0 & (x>\ell/2)
    \end{cases},
\end{align}
with $E<V_0$.
The storage bottle produces a different potential for antineutrons:
\begin{align}
    U_\nbar(x)=
    \begin{cases}
        V_0'-iW_0' & (x<-\ell/2)\\
        0 & (-\ell/2<x<\ell/2)\\
        V_0'-iW_0' & (x>\ell/2)
    \end{cases}.
\end{align}

In a semi-classical picture, a neutron starts at $x=-\ell/2$ at $t=0$ and moves with constant velocity $v=\sqrt{2E/m}$ between the walls, reflecting at intervals of $T=\ell/v$. This oscillatory motion corresponds to the spatial part of the wavefunction $\exp(\pm ikx)$, where the sign denotes the direction of motion.

Next, we consider the effect of reflections at $x=\pm \ell/2$ on the neutron and antineutron wavefunctions
to relate the change of the wavefunctions before and after a reflection.
Using $k=\sqrt{2mE}/\hbar$, $k'=\sqrt{2m(E-V_0)}/\hbar$, and $k''=\sqrt{2m[E-(V_0'-iW_0')]}/\hbar$, 
we define the following parameters:
\begin{align}
    R_\nbar&=\left|\frac{k-k''}{k+k''}\right|,\label{eq:Rbar}\\
    \varphi_n&=\arg \frac{k-k'}{k+k'},\\
    \varphi_\nbar&=\arg\frac{k-k''}{k+k''}\label{eq:varphi}.
\end{align}
The reflection coefficients for the neutron and antineutron are expressed as $\exp(i\varphi_n)$ and $R_\nbar\exp(i\varphi_\nbar)$, respectively.  The antineutron reflectivity is given by $R_\nbar^2$.

The wavefunctions of the neutron and antineutron for $(N-1)T<t<NT$, i.e. between $(N-1)$-th and $N$-th reflections, are represented by $\psi_n^{(N)}(t)$ and $\psi_\nbar^{(N)}(t)$, respectively.
The wavefunctions are subject to change by the $N$-th reflection as
\begin{align}
    \psi_n^{(N+1)}(NT)&=\exp(i\varphi_n)\psi_n^{(N)}(NT),\label{eq_psi_n_orig}\\
    \psi_\nbar^{(N+1)}(NT)&=R_\nbar\exp(i\varphi_\nbar)\psi_\nbar^{(N)}(NT).\label{eq_psi_nbar_orig}
\end{align}
Since we are interested in the relative phase between the neutron and the antineutron wavefunctions, we introduce the relative phase shift $\Delta\varphi=\varphi_\nbar-\varphi_n$, and omit the common phase factor $\exp(i\varphi_n)$ in wavefunctions,
by redefining
\begin{align}
[\exp(i\varphi_n)]^{N-1}\psi_{n}^{(N)}(t)&\to \psi_{n}^{(N)}(t),\\
[\exp(i\varphi_n)]^{N-1}\psi_{\nbar}^{(N)}(t)&\to \psi_{\nbar}^{(N)}(t),
\end{align}
for $(N-1)T<t<NT$.
We also introduce a complex parameter  $\tilde{R}=R_\nbar\exp(i\Delta\varphi)$ to further simplify the notations.
Hence, Eqs.~(\ref{eq_psi_n_orig}) and (\ref{eq_psi_nbar_orig}) are rewritten as
\begin{align}
    \psi_n^{(N+1)}(NT)&=\psi_n^{(N)}(NT)\label{eq_psi_n},\\
    \psi_\nbar^{(N+1)}(NT)&=\tilde{R}\psi_\nbar^{(N)}(NT).\label{eq_psi_nbar}    
\end{align}
As Eq.~(\ref{eq_psi_n}) shows that the neutron wavefunction is unchanged by a reflection, it justifies the use of Eq.~(\ref{common_psi_n}) for $\psi_n^{(N)}(t)$ and therefore Eq.~(\ref{eq:Schrodinger}) for $\psi_\nbar^{(N)}(t)$. 

\subsection{Time evolution of the antineutron wavefunction}
Now, we will use Eq.~(\ref{eq_psi_nbar}) to account for the effects of the wall reflections and Eq.~(\ref{eq:Schrodinger}) for time evolution between the reflections, and derive the time evolution of the antineutron wavefunction. 

The Schr\"{o}dinger equation for $\psi_\nbar^{(N)}$ (Eq.~(\ref{eq:Schrodinger})) has a solution similar to Eq.~(\ref{eq:psi_nbar_sol}):
\begin{align}
    \psi_\nbar^{(N)}(t)&=-\frac{\delta m}{2\Delta E}\exp\left[-\frac{i(E-\Delta E)t}{\hbar}\right]+a^{(N)}\exp\left[-\frac{i(E+\Delta E)t}{\hbar}\right],\label{psi_nbar_N}
\end{align}
where $a^{(N)}$ is a constant.
A recurrence relation between $a^{(N)}$ and $a^{(N+1)}$ is obtained by using Eq.~(\ref{eq_psi_nbar}) as 
\begin{align}
    &-\frac{\delta m}{2\Delta E}\exp\left(\frac{i\Delta E\,NT}{\hbar}\right)+a^{(N+1)}\exp\left(\frac{-i\Delta E\,NT}{\hbar}\right)\nonumber\\
    =&-\frac{\delta m}{2\Delta E}\tilde{R}\exp\left(\frac{i\Delta E\,NT}{\hbar}\right)+\tilde{R}a^{(N)}\exp\left(\frac{-i\Delta E\,NT}{\hbar}\right),
\end{align}
leading to
\begin{align}
    a^{(N+1)}=\tilde{R}a^{(N)}+\frac{\delta m}{2\Delta E}(1-\tilde{R})\exp\left(\frac{2i\Delta E\,NT}{\hbar}\right).\label{reccurence}
\end{align}
As $a^{(1)}=\delta m/(2\Delta E)$ from Eq.~(\ref{eq:psi_nbar_sol}), the series of $a^{(N)}$ can be determined by repeatedly using the recurrence relation:
\begin{align}
a^{(N)}&=\frac{\delta m}{2\Delta E}\left[1-\frac{(1-\tilde{R}^N)(1-e^{2i\Delta E\,T/\hbar})-(1-\tilde{R})(1-e^{2i\Delta E\,NT/\hbar})}{\tilde{R}-e^{2i\Delta E\,T/\hbar}}\right]\label{eq:aN_exact_solution}\\
&=\frac{\delta m}{2\Delta E}\left[\left(\exp\frac{2i\Delta E\,T}{\hbar}-1\right)\frac{e^{2i\Delta E\,NT/\hbar}-\tilde{R}^N}{\tilde{R}-e^{2i\Delta E\,T/\hbar}}+\exp\frac{2i\Delta E\,NT}{\hbar}\right],\\
\intertext{
and if we assume $\Delta E\,T\ll \hbar$, the quasi-free condition for the time between reflections,
}
&\approx \frac{\delta m}{2\Delta E}\left(-\frac{2i\Delta E\,T}{\hbar}\frac{e^{2i\Delta E\,NT/\hbar}-\tilde{R}^N}{e^{2i\Delta E\,T/\hbar}-\tilde{R}}+\exp\frac{2i\Delta E\,NT}{\hbar}\right)\\
&=-i\frac{T}{\tau_\nnbar}\frac{e^{2i\Delta E\,NT/\hbar}-\tilde{R}^N}{e^{2i\Delta E\,T/\hbar}-\tilde{R}}+\frac{\delta m}{2\Delta E}\exp\frac{2i\Delta E\,NT}{\hbar}.\label{aN_sol3}
\end{align}
Substituting Eq.~(\ref{aN_sol3}) into Eq.~(\ref{psi_nbar_N}), we obtain 
\begin{align}
    \psi_\nbar^{(N)}(NT)=-i\frac{T}{\tau_\nnbar}\frac{e^{2i\Delta E\,NT/\hbar}-\tilde{R}^N}{e^{2i\Delta E\,T/\hbar}-\tilde{R}}\exp\left[-\frac{i(E+\Delta E)NT}{\hbar}\right],
\end{align}
which gives the antineutron appearance probability at $t=NT$ (just before the $N$-th reflection) 
\begin{align}
    P_\nbar(NT)=\frac{T^2}{\tau_\nnbar^2}\left|\frac{e^{2i\Delta E\,NT/\hbar}-\tilde{R}^N}{e^{2i\Delta E\,T/\hbar}-\tilde{R}}\right|^2.\label{nbar_appearance_prob2}
\end{align}
If the quasi-free condition, $\Delta E\,t\ll \hbar$, is met for the elapsed time $t=NT$, the probability approaches $(NT)^2/\tau_\nnbar^2=t^2/\tau_\nnbar^2$ in the limit $\tilde{R}\to 1$, consistent with Eq.~(\ref{nbar_appearance_prob}).
Otherwise, for a sufficiently large $N$, the probability converges to a constant:
\begin{align}
P_\nbar(NT)\xrightarrow{N\to \infty} \frac{T^2}{\tau_\nnbar^2}\frac{1}{|e^{2i\Delta E\,T/\hbar}-\tilde{R}|^2}.\label{nbar_appearance_prob2_largeN}
\end{align}

\begin{figure*}[t]
\centerline{
\subfloat[$\Delta E\,T=0.01\hbar$.]{\includegraphics[scale=0.55]{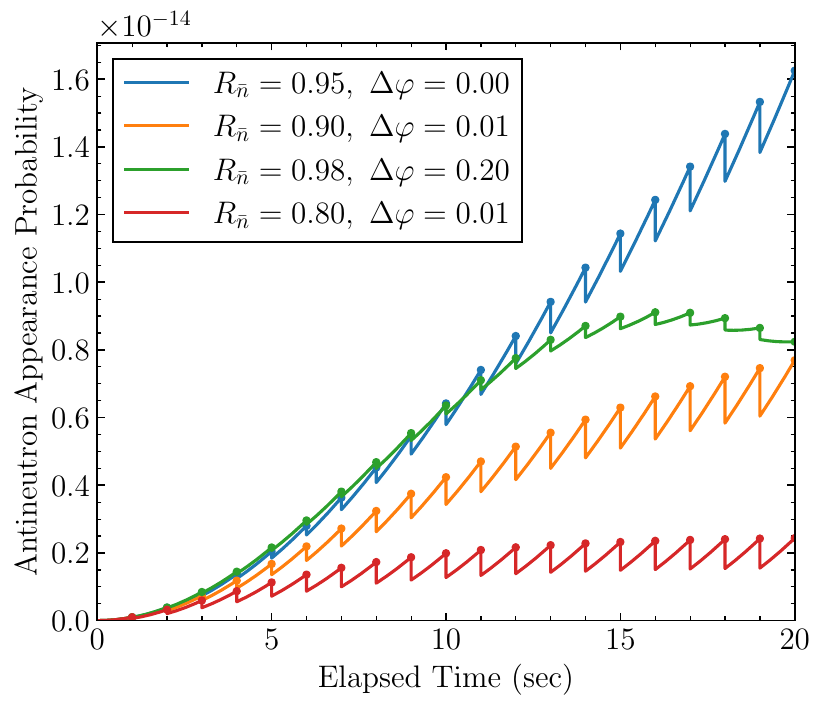}%
}
\hfil
\subfloat[$\tilde{R}=0.95$.]{\includegraphics[scale=0.55]{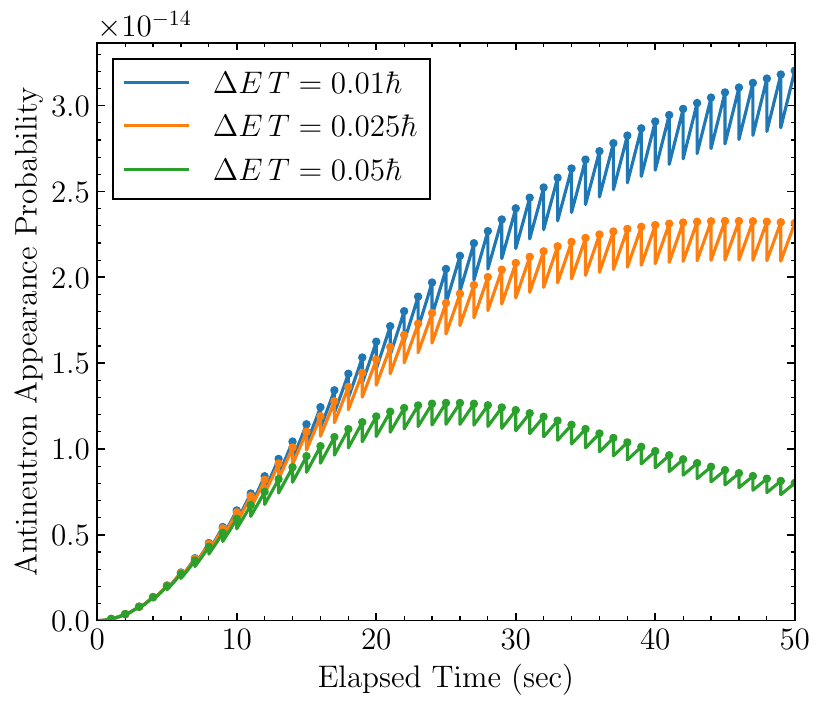}%
}
}
\caption{Antineutron appearance probability $P_\nbar(t)=|\psi_\nbar(t)|^2$ for different conditions of $\Delta E$ and $\tilde{R}=R_\nbar\exp(i\Delta\varphi)$. The lines show numerical results using the Runge--Kutta method, and the dots represent values from Eq.~(\ref{nbar_appearance_prob2}). We set $T=1\,\mathrm{sec}$ and $\delta m=\hbar/(10^8\,\mathrm{sec})$.}
\label{fig_timeevolution}
\end{figure*}

To validate this result, we compare the numerical solution of the Schr\"{o}dinger equation (Eq.~(\ref{eq:Schrodinger})), calculated using the Runge–Kutta method, with the  probability given by Eq.~(\ref{nbar_appearance_prob2}).
As shown in Fig.~\ref{fig_timeevolution}, the two approaches give the same probabilities at multiples of $T$.

\subsection{Antineutron annihilation rate}
From an experimental point of view, antineutrons in a storage bottle are identified via their annihilation with nucleons.
Therefore, the Figure of Merit (FoM) of the $\nnbar$ oscillation experiment using UCNs in a storage bottle should be taken to be a quantity related not to the antineutron appearance probability but to the antineutron annihilation rate.
As the annihilation probability for an antineutron incident on the wall is given by $1-R_\nbar^2$, the annihilation rate is 
\begin{align}
    P^{\mathrm{(ann.)}}(t=NT)&=\sum_{j=1}^N (1-R_\nbar^2)P_\nbar(jT),
\intertext{and except for the case $R_\nbar\approx 1$,}
    &\xrightarrow{N\to\infty}\frac{NT^2}{\tau_\nnbar^2}\frac{1-R_\nbar^2}{|e^{2i\Delta E\,T/\hbar}-\tilde{R}|^2}+\text{const.}\label{Pann_largeN}
\end{align}
Since the annihilation probability in $N$ independent measurements of a free neutron of flight time $T$ is given by $NT^2/\tau_{\nnbar}^2$ from Eq.~(\ref{nbar_appearance_prob}), 
we define 
\begin{align}
\mathrm{FoM}= \frac{1-R_\nbar^2}{|e^{2i\Delta E\,T/\hbar}-\tilde{R}|^2},
\end{align}
the factor that accounts for the enhancement of the annihilation probability due to the confinement of UCNs in a storage bottle.
The experimental sensitivity to the oscillation time $\tau_\nnbar$ improves by a factor of $\sqrt{\text{FoM}}$.
In the limits of $\Delta \varphi\to 0$ and $\Delta E\to 0$, the FoM becomes
\begin{align}
    \frac{1-R_\nbar^2}{|e^{2i\Delta E\,T/\hbar}-\tilde{R}|^2}\xrightarrow{\substack{\Delta \varphi\to 0\\ \Delta E\to 0}}\frac{1+R_\nbar}{1-R_\nbar},
\end{align}
which is an increasing function of the antineutron reflectivity $R_\nbar^2$.

\begin{figure*}[t]
\centering
\includegraphics[scale=0.55]{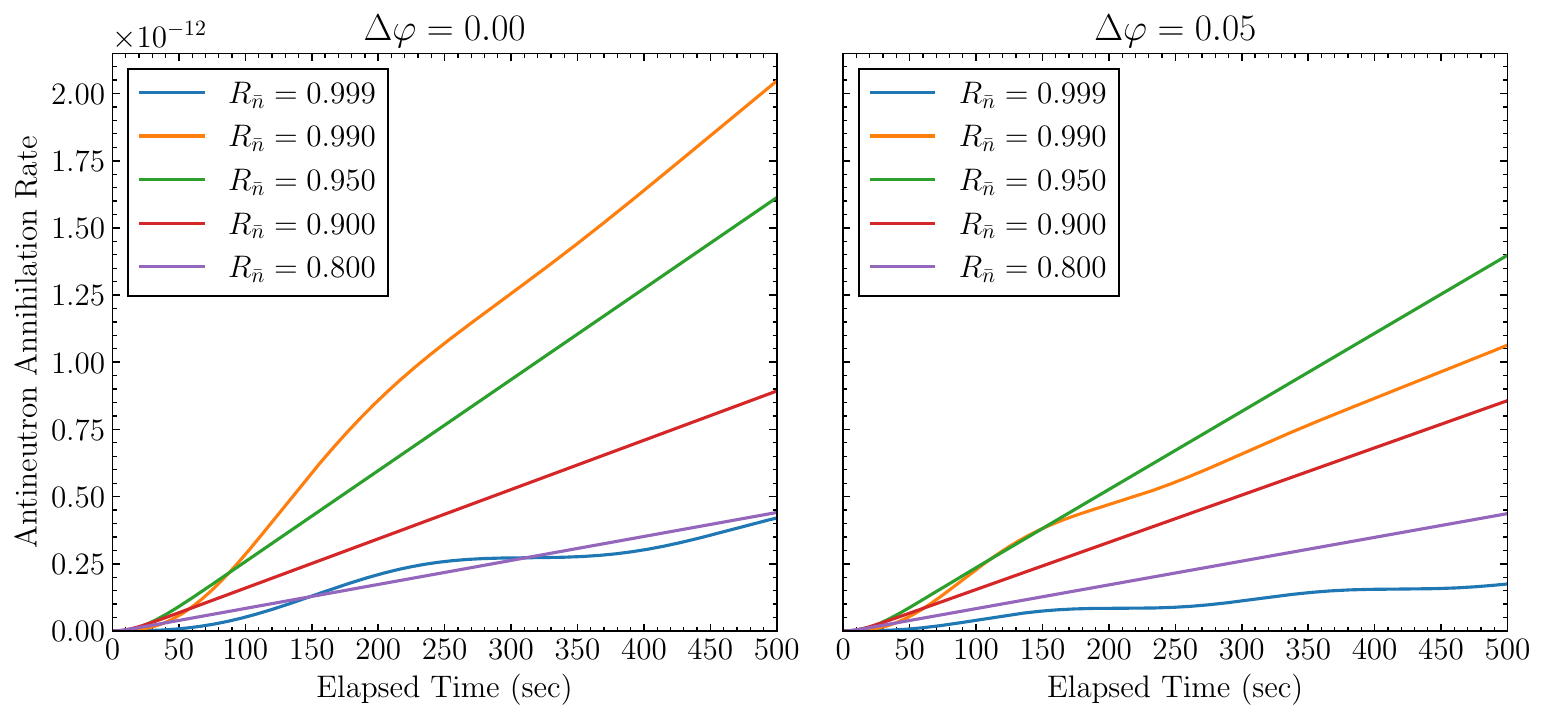}
\caption{Antineutron annihilation rate for different conditions of $\tilde{R}=R_\nbar\exp(i\Delta\varphi)$. We set $T=1\,\mathrm{sec}$, $\delta m=\hbar/(10^8\,\mathrm{sec})$, and $\Delta E\,T=0.01\hbar$.}
\label{fig_annihilationrate}
\end{figure*}

Figure~\ref{fig_annihilationrate} shows the FoM calculated for $R_\nbar=0.999$, 0.99, 0.95, 0.9, 0.8 and $\Delta\varphi=0$, 0.05.
When $R_\nbar$ is very close to unity, a small absorption probability $1-R_\nbar^2$ suppresses the annihilation rate in this time range.  
Moreover, the annihilation rate has a strong $\Delta \varphi$ dependence when $1-R_\nbar\lesssim 2\Delta E\,T/\hbar$, due to the denominator on the rhs of Eq.~(\ref{nbar_appearance_prob2}).
In contrast, for realistic conditions of the antineutron reflectivity around 0.8--0.9 (i.e., $R_\nbar$ around 0.9--0.95), the slope is almost proportional to $(1+R_\nbar)/(1-R_\nbar)$, ignoring moderate dependence on $\Delta\varphi$ and $\Delta E$.

The authors of Ref.~\cite{Kerbikov2004} treated the neutron--antineutron oscillation of trapped UCNs by the same approach as ours, but under an assumption of an idealized magnetic shield, where $\Delta E=|\mu_n|B=0$.
They obtained the antineutron wavefunction and the annihilation rate (Eqs.~(56) and (60) of Ref.~\cite{Kerbikov2004}) as functions of $R_\nbar$ and $\Delta\varphi$ ($\rho$ and $\theta$ in their notation).
This corresponds to our results in Eqs. (\ref{nbar_appearance_prob2}) and (\ref{Pann_largeN}) in the limit of zero magnetic field.
It is worth noting that they regarded the case of $\Delta \varphi=0$ as an ideal but unrealistic situation.
However, we demonstrate in the next section that this is indeed feasible.

\section{Considerations on a UCN reflecting material}\label{sec:UCNmaterial}
In general, the pseudopotential is proportional to the scattering length, as introduced in Section~\ref{sec:UCNreflection}.
The neutron scattering lengths are well determined for most nuclides~\cite{Sears1992}, whereas those of the antineutrons are not directly determined.
Instead, previous results of X-ray spectroscopy of antiprotonic atoms suggest
\begin{align}
    a_{\nbar A}=(1.54\pm 0.03)\cdot A^{0.311\pm 0.005}-i(1.00\pm 0.04)\,\mathrm{fm},\label{b_nbarA}
\end{align}
assuming $\nbar A$ and $\bar{p}A$ strong interactions are equivalent~\cite{Batty2001-hh}.
The real part of Eq.~(\ref{b_nbarA}), scaling roughly  $A^{1/3}$, reflects the nuclear radius. The finite imaginary part stems from the diffuseness of the nuclear potential~\cite{Batty1983-sn}.
The outcomes of this model have  been used in some previous studies, for example in Table I of Ref.~\cite{Nesvizhevsky2019-vw}.
To obtain a large reflectivity for a wide range of kinetic energies,
a large real part and a small imaginary part are desirable.
Given the limited knowledge of $U_\nbar$, we investigate the impact of the pseudopotentials $U_n$ and $U_\nbar$ on the sensitivity of a $\nnbar$ oscillation search, considering values around $U_\nbar = 100 - 10i\,\mathrm{neV}$.

\begin{figure*}[t]
\centerline{
\subfloat[Reflectivity $R^2$.]{\includegraphics[scale=0.55]{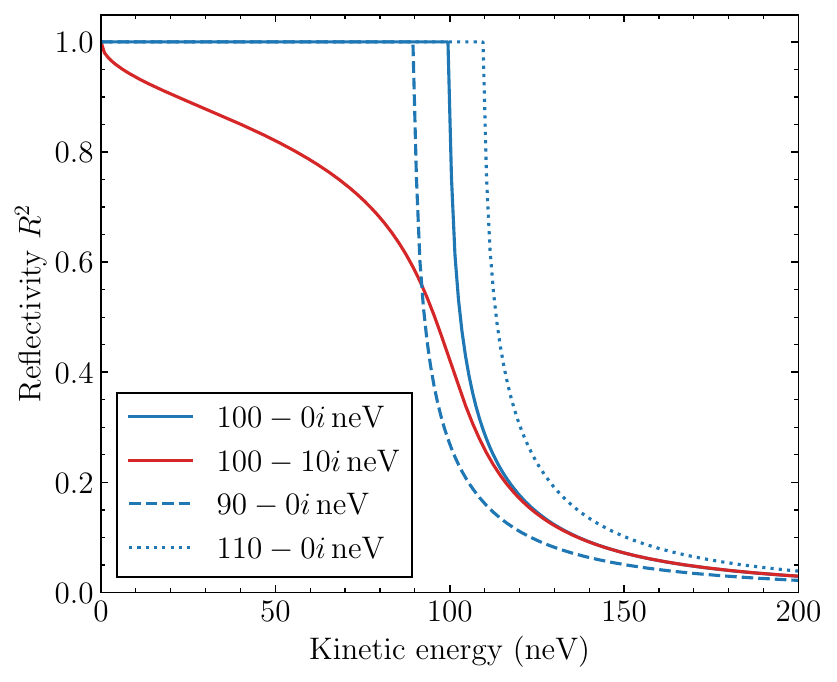}%
}

\hfil
\subfloat[Phase shift $\varphi$.]{\includegraphics[scale=0.55]{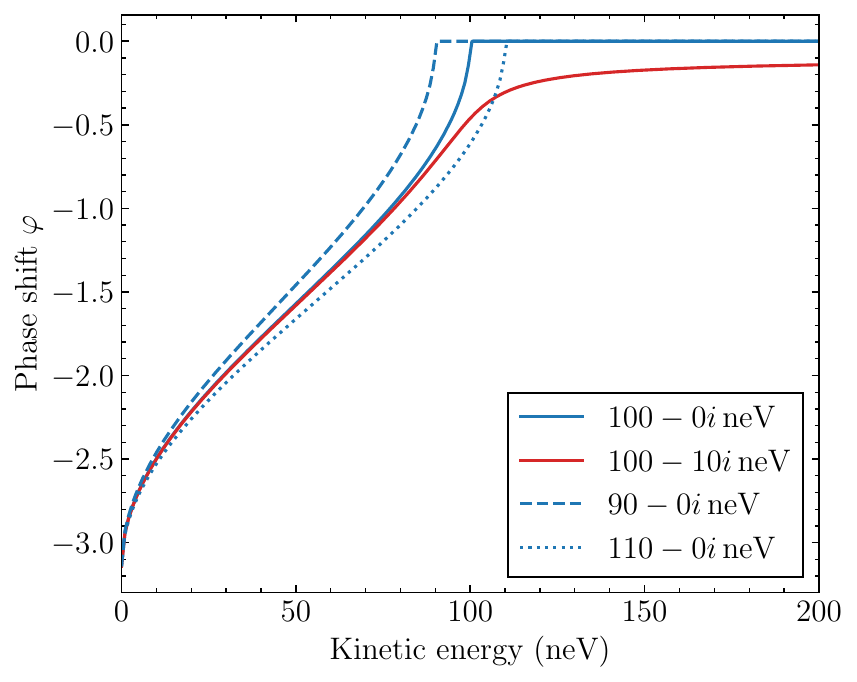}%
}
}
\caption{The reflectivity and phase shift caused by a pseudopotential of $V_0-iW_0$.}
\label{fig_reflection}
\end{figure*}

Figure~\ref{fig_reflection} shows the calculated reflectivity and phase shift using Eqs.~(\ref{eq:Rbar})--(\ref{eq:varphi}) for the pseudopotential $U_\nbar$.
For reference, those for real pseudopotentials, $90\text{--}110\,\mathrm{neV}$, are also shown.
It should be emphasized that, despite the large antineutron annihilation cross section, a reflectivity exceeding 0.8 can still be achieved for kinetic energies below $50\,\mathrm{neV}$.
Another important remark is that the phase shift for $U_\nbar=100-10i\,\mathrm{neV}$ almost coincides with that for $U_n=100\,\mathrm{neV}$; the difference is less than $0.01\,\mathrm{rad}$ below $50\,\mathrm{neV}$.
This is the case only when the real parts of the pseudopotentials are close to each other.
A very small $|\Delta \varphi|$ is favored in enhancing the annihilation rate, as demonstrated in Fig.~\ref{fig_annihilationrate}.

It should be noted that the retardation time in a wave packet formalism, given by $\hbar (d\varphi/dE)$~\cite{Kerbikov2019-lb}, is also comparable between neutrons and antineutrons; the spacial distance between the two wave packets after a reflection is of the order of 
$1\,\mathrm{nm}$, much smaller than the de Broglie wavelength of UCNs.
This also reinforces the validity of the semi-classical treatment~\cite{Kerbikov2003-cw}.

\begin{figure*}[t]
\centering
\includegraphics[scale=0.55]{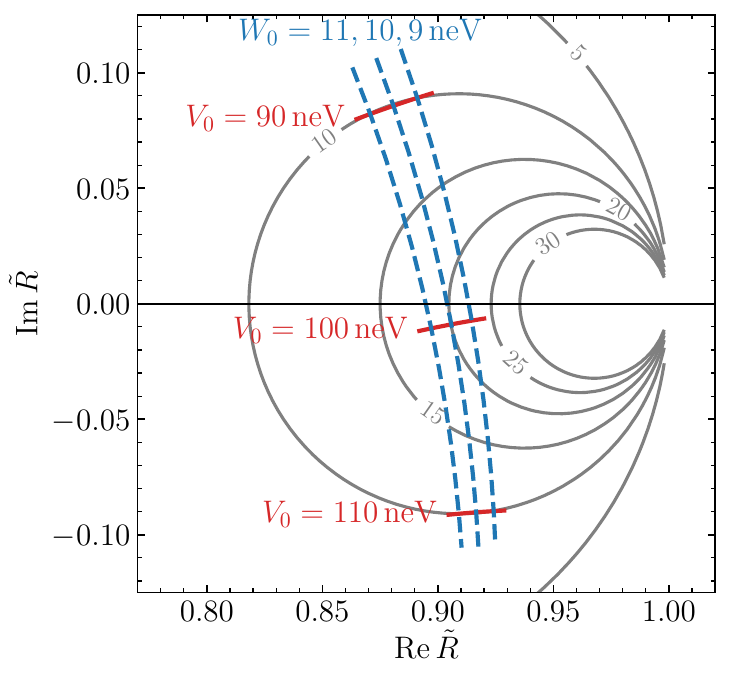}%
\caption{Dependence of $\tilde{R}=R_\nbar\exp(i\Delta\varphi)$ on the antineutron pseudopotential strength $V_0-iW_0$ for $50\,\mathrm{neV}$ UCNs reflecting from a material with $U_n=100\,\mathrm{neV}$. A contour plot of the FoM in case of $\Delta E=0$ is overlaid in gray.}
\label{fig_Rtilde}
\end{figure*}

Next, to quantitatively investigate the deterioration of the FoM caused by the difference between $U_n$ and $\mathrm{Re}\,U_\nbar$,
we consider a material with $U_n=100\,\mathrm{neV}$, and we vary the real and imaginary parts of $U_\nbar=V_0-iW_0$ within 10\% around the central value of $100-10i\,\mathrm{neV}$.
Figure~\ref{fig_Rtilde} shows the trajectories of $\tilde{R}=R_\nbar\exp(i\Delta\varphi)$ for UCNs with a kinetic energy of $50\,\mathrm{neV}$, as a function of $V_0$ and $W_0$ with either $V_0$ or $W_0$ held fixed.
A contour plot of the FoM in the $\tilde{R}$ plane is overlaid in Figure~\ref{fig_Rtilde} for a case of $\Delta E=0$ . 
It is found that the FoM depends more strongly on $V_0$ than on $W_0$.
The FoM for the baseline pseudopotential $V_0-iW_0=100-10i\,\mathrm{neV}$ is 20.
Whereas a $\pm 10\%$ variation in $W_0$ results in only a $\mp 10\%$ change in the FoM, the same $\pm 10\%$ variation in $V_0$ halves the FoM 
to 10.

Recalling that the antineutron pseudopotential is proportional to the antineutron--nucleus scattering length, the uncertainty in the scattering length, in particular its real part, strongly affects both the discovery potential of the experiment and the uncertainty of the derived upper limit on the oscillation time $\tau_\nnbar$.
Therefore, this uncertainty must be carefully assessed in designing the UCN storage bottle. However, Eq.~(\ref{b_nbarA}) does not provide uncertainties for individual nuclides; it only constrains the overall trend of scattering lengths as a function of mass number $A$.

\section{Toward refined determinations of antineutron scattering lengths}\label{sec:scatteringlength}
Whereas neutron scattering length has been determined using neutron interferometers~\cite{Fujiie2024-zw},
the same technique cannot be applied for antineutrons.
The primary reason is that antineutrons, which were produced by the charge-exchange reaction ($\bar{p}+p\to \bar{n}+n$), had kinetic energies in the MeV range or above~\cite{Bressani2003-lc}, compared to those of thermal neutrons around $0.025\,\mathrm{eV}$.
Instead, spectroscopic studies of antiprotonic atoms, in which a negatively charged antiproton is bound to a nucleus, provide insight into the strong interaction between the antiproton and the nucleus~\cite{Batty1997-nd,Friedman2007-vu,Friedman2014-ry}.
Charge symmetry ($p\leftrightarrow n$ and $\bar{p}\leftrightarrow\bar{n}$) relates the antiproton--nucleus interaction and the antineutron--nucleus interaction.

The low-energy antiproton--nucleus strong interaction is described by an optical potential $V(r)$, commonly assumed to be proportional to the nucleon density $\rho(r)$:
\begin{align}
    V(r)=-\frac{2\pi}{\mu}\left(1+\frac{\mu}{M_N}\right)b_0\rho(r),
\end{align}
where $\mu$ is the reduced mass, $M_N$ is the nucleon mass, and  $b_0$ is a complex parameter related to the $s$-wave antiproton--nucleon  scattering length.
A global fit to X-ray data of antiprotonic atoms across the periodic table was performed to determine $b_0$ as well as proton and neutron density distributions which were assumed to follow two-parameter Fermi distributions~\cite{Friedman2005-yl}.
The aforementioned antineutron scattering length in Eq.~(\ref{b_nbarA}) was derived by solving the Schr\"{o}dinger equation for a low-energy antineutron scattering off the optical potential, which was obtained for each species of antiprotonic atom~\cite{Batty2001-hh}.

Recently, it was pointed out that this type of optical potential cannot reproduce the energy levels of antiprotonic $\ce{^40Ca}$ and $\ce{^48Ca}$ atoms without the inclusion of an isovector term proportional to $\delta\rho(r)=\rho_n(r)-\rho_p(r)$, where $\rho_{n/p}(r)$ represents the neutron/proton density distribution~\cite{yoshimura2025}. 
Moreover, the authors of Ref.~\cite{yoshimura2025} showed that the energy levels strongly depends on nucleon density distributions, indicating that the optical potential parameters such as $b_0$, should be determined using a realistic distribution rather than of a simple two-parameter Fermi model.

This situation calls for precise measurements of antiprotonic atoms with careful analyses based on the current understanding of nuclear structures.
In Ref.~\cite{Higuchi:2025YA}, a new experiment employing  multi-pixel transition-edge sensor (TES) superconducting microcalorimeters is proposed to improve the antiproton-nucleus optical potential model. The TES detector offers an order of magnitude higher energy resolution than high-purity germanium detectors used in previous experiments~\cite{Trzcinska2009-xo}.

Another method for determining the scattering length is the measurement of antineutron--nucleus cross sections in the low energy regime.
For sufficiently low energies, where higher-order partial waves beyond the $s$-wave are negligible, the elastic scattering and annihilation cross sections, denoted by $\sigma_\text{el}$ and $\sigma_\text{ann}$, respectively, can be expanded as a series in the wavenumber of the incident antineutrons $k$:
\begin{align}
\sigma_\mathrm{el}&=4\pi|a|^2(1-2a_\mathrm{I}k),\\
\sigma_\mathrm{ann}&=\frac{4\pi}{k}a_\mathrm{I}-8\pi a_\mathrm{I}^2,
\end{align}
where $a=a_\mathrm{R}-ia_\mathrm{I}$ is the scattering length.
This measurement requires an antineutron beam with unprecedented low kinetic energy, as low as $\sim 0.05\,\mathrm{MeV}$ ($\sim 10\,\mathrm{MeV}/c$ in momentum), which can be produced at backward angles via the charge-exchange reaction. Antiproton beams suitable for such experiments are potentially available at the Antiproton Decelerator (AD) at CERN~\cite{Amsler:2930906}.
Such experiments also have implications for various areas of hadron physics.
These physics cases are discussed in Ref.~\cite{Filippi2025}.

These hadron-physics approaches will enable a rigorous evaluation of antineutron pseudopotentials, and provide essential input for $\nnbar$ oscillation experiments,  using not only the UCN method, but also other approaches~\cite{Nesvizhevsky2019-vw,Gudkov2020-dn,Protasov2020-gy}.
\section{Conclusion}\label{sec:conclusion}
We investigated neutron--antineutron oscillations in a one-dimensional trap potential using the semi-classical treatment, and derived the antineutron wavefunction and the antineutron annihilation rate assuming the quasi-free condition for the time between the reflections 
 $\Delta E\,T\ll \hbar$. The stricter condition for the total elapsed time $\Delta E \,t \ll\hbar$ is not  required for this derivation.
Based on the linear dependence of the antineutron annihilation rate on the elapsed time, 
we defined the experimental FoM as a function of $\tilde{R}=R_\nbar\exp(i\Delta \varphi)$ and $\Delta E$.
The complex parameter $\tilde{R}$ can be calculated using the neutron and antineutron pseudopotentials of the reflecting material of a UCN storage bottle.

This model was used to study the dependence of the FoM on the real and imaginary parts of the antineutron pseudopotential.
To enhance the FoM, we found that the real parts of these pseudopotentials should be close to each other, which minimizes the relative phase shift $|\Delta\varphi|$ even for a finite imaginary part.
This implies that both the real part of the antineutron pseudopotential and its uncertainty should be well understood.

Whereas the neutron pseudopotential is obtained from the well-known neutron scattering length, the antineutron scattering length for each element needs to be evaluated in dedicated experiments.
To this end, we proposed two approaches: precision X-ray spectroscopy of antiprotonic atoms and low-energy antineutron scattering.

Finally, we point out that this approach can be generalized to a three-dimensional storage bottle, where UCNs follow ballistic trajectories under gravity and undergo non-periodic and inclined reflections from the walls.
Since the reflection coefficients, $\exp(i\varphi_n)$ and $R_{\nbar}\exp(i\varphi_\nbar)$, depend on the normal components of the wavevectors,
the optimization principle for the reflecting material outlined above remains applicable.
Once the material is optimized, detailed numerical simulations can be used to refine realistic experimental conditions for an $\nnbar$ oscillation search, accounting for non-ideal effects, such as surface roughness and magnetic-field inhomogeneities.  

\section*{Acknowledgements}

The authors would like to acknowledge S.~Kawasaki, M.~Kitaguchi, K.~Mishima, T.~Okudaira, T.~Shima, and H.~M.~Shimizu for valuable discussions. 
T.H.~acknowledges the support of the JST FOREST Program (Grant No.~JPMJFR2237).


%



\let\doi\relax


\bibliographystyle{ptephy}
\bibliography{reference}


\end{document}